\documentclass[useAMS,usenatbib,usegraphicx]{mn2e}

\newcommand\lsim{\mathrel{\rlap{\lower4pt\hbox{\hskip1pt$\sim$}}
        \raise1pt\hbox{$<$}}}
\newcommand\gsim{\mathrel{\rlap{\lower4pt\hbox{\hskip1pt$\sim$}}
        \raise1pt\hbox{$>$}}}

\newcommand{\br}{\mathrm {b}}

\newcommand{\tot}{\mathrm{tot}}

\title{Gas in Simulations of High Redshift Galaxies and Minihalos}
\author[Smadar Naoz, Rennan Barkana \& Andrei Mesinger]
{Smadar Naoz$^1$\thanks{E-mail: smadar@wise.tau.ac.il},
Rennan Barkana$^{1,2}$
\& Andrei Mesinger$^3$\thanks{Hubble Fellow}\\
$^{1}$Raymond and Beverly Sackler School of Physics and Astronomy, Tel Aviv
University, Tel Aviv 69978, Israel \\
$^{2}$ Division of Physics, Mathematics and Astronomy, California Institute of Technology, Mail Code 130-33, Pasadena, CA 91125, USA \\
$^{3}$ Department of Astrophysical Sciences, Princeton University, Princeton, NJ 08544, USA}

\begin{document}
\pagerange{\pageref{firstpage}--\pageref{lastpage}} \pubyear{2009}
\maketitle
\label{firstpage}

\begin{abstract}
  We study the gas content of halos in the early universe using high
  resolution hydrodynamical simulations. We extract from the
  simulations and also predict based on linear theory the halo mass
  for which the enclosed baryon fraction equals half of the mean
  cosmic fraction. We find a rough agreement between the simulations
  and the predictions, which suggests that during the high-redshift
  era before stellar heating, the minimum mass needed for a minihalo
  to keep most of its baryons throughout its formation was $\sim 3
  \times 10^4$~M$_\odot$.  We also carry out a detailed resolution
  analysis and show that in order to determine a halo's gas fraction
  even to $20\%$ accuracy the halo must be resolved into at least 500
  dark matter particles.
\end{abstract}


\section{Introduction}\label{intro}

The formation of galaxies is one of the most important research areas
in cosmology. Within the simplified hierarchical scenario of a
universe governed by a cosmological constant and cold dark matter, the
density profiles of forming halos have been well characterized
\citep{NFW}. However, the complex processes of gas dynamics, such
as cooling and heating mechanisms, that are responsible for the
formation of luminous objects still pose many theoretical
difficulties.

Numerical calculations show that the first generation of galaxies
formed at very high redshifts inside collapsing halos \citep[starting
at $z \sim 65$;][]{NNB}, corresponding to high peaks of the primordial
density field. Indeed, The {\it Wilkinson Microwave Anisotropy Probe}
({\it WMAP}) measured a Thomson scattering optical depth of $\tau_e =
0.09\pm0.019$
from their 5-year data \citep{Dunkley08}. When combined with simple
analytic perscriptions of the growth of the global ionized fraction
\citep[e.g., fig. 22][fig. 9]{rev,MJH06}, this measurement suggests
that reionization began at very high redshifts, $z\gsim15$.
 This means that a high enough abundance of luminous
objects must have existed at that time, since these first luminous
objects are expected to have heated and reionized their surroundings
\citep[e.g.,][]{rev,WL03,HH03,Cen}. The formation of a luminous
object inside a halo requires, of course, gas to be inside the halo.
Even in halos that are too small for cooling via atomic hydrogen,
i.e., minihalos, the gas content can have substantial, and observable,
astrophysical effects. In addition to the posibility of hosting
astrophysical sources, minihalos may produce a 21-cm signature
(\citet{Kuhlen,Shapiro06,NB08} but see \citet{Furlanetto06}), and they can
block ionizing radiation and produce an overall delay in the global
progress of reionization \citep[e.g.,][]{bl02, iliev2, iss05,
  mcquinn07}.  Thus, the evolution of the halo gas fraction at various
epochs of the universe is of prime importance, particularly in the
early universe.

The estimation of the gas fraction in simulations and
semi-analytical models has been extensively investigated and used
for various purposes
\citep[e.g.,][]{Efstathiou,Shapiro94,TW96,Quinn,HG97,Bromm99,gnedin00,KI00,Abel02,Bromm02,Helly,BL04,Dijkstra,Oshea05,Andrei,NB07,MBH08,Trenti}.
When investigating this issue with simulations, a large volume is
needed for an adequate statistical sample of halos, but on the other
hand it is critical to maintain the proper resolution. The results of
\citet{SH03} showed that in order to determine the mass and merger
history of each halo even crudely in simulations, each halo must be
resolved into 500 particles. In the high redshift regime, where halos
are small and rare, these resolution requirements are not easy to
achieve. Nonetheless, it is important to do so in order to understand
the formation of the first objects.

Consider the various scales involved in the formation of non-linear
objects containing DM and gas. On large scales (small wavenumbers)
gravity dominates halo formation and gas pressure can be neglected. On
small scales, on the other hand, the pressure dominates gravity and
prevents baryon density fluctuations from growing together with the
dark matter fluctuations. The relative force balance at a given time
can be characterized by the \citet{jeans} scale, which is the minimum
scale on which a small gas perturbation will grow due to gravity
overcoming the pressure gradient. As long as the Compton scattering of
the cosmic microwave background (CMB) on the residual free electrons
after cosmic recombination kept the gas temperature coupled to that of
the CMB, the Jeans mass was constant in time. However, at $z\sim 200$
the gas temperature decoupled from the CMB temperature and the Jeans
scale began to decrease with time as the gas cooled adiabatically. Any
overdensity on a scale more massive than the Jeans mass at a given
time can begin to collapse, due to a lack of sufficient
pressure. However, the Jeans mass is related only to the evolution of
perturbations at a given time. When the Jeans mass itself varies with
time, the overall suppression of the growth of perturbations depends
on a time-averaged Jeans mass.

\citet{cs} defined a ``filtering mass'' that describes the highest
mass scale on which the baryonic pressure still manages to suppress
the linear baryonic fluctuations significantly. \citet{gnedin00}
suggested, based on a simulation, that the filtering mass also
describes the largest halo mass whose gas content is significantly
suppressed compared to the cosmic baryon fraction. The latter mass
scale, in general termed the ``characteristic mass'', is defined as
the halo mass for which the enclosed baryon fraction equals half the
mean cosmic fraction. Thus, the characteristic mass distinguishes
between gas-rich and gas-poor halos. Many semi-analytical models of
dwarfs galaxies often use the characteristic mass scale in order to
estimate the gas fraction in halos
\citep{Bullock,Benson02a,Benson02b,Somerville}.Recently,
\citet{Hoeft} and \citet{Okamoto} showed that the characteristic
mass scale does not agree with the \citet{cs} filtering mass in the
low-redshift, post-reionization regime.

In this paper we explore the very high-redshift regime using
three-dimensional hydrodynamical simulations based on \citet{Andrei}.
 They investigated the effects of a photo-ionizing ultraviolet (UV) flux on the
collapse and cooling of small halos in relic HII regions at high redshift, by varying the strength and duration of a transient UV and persistent Lyman-Werner (LW) background.
 We consider two
different scenarios presented there (see Section~\ref{sec:sims} for a
description of the simulations). We perform a resolution study and
place a lower limit on the number of particles needed in a simulated
halo in order to accurately determine the gas fraction in halos
(Section~\ref{sec:res}). We also compare the numerical evolution of
the filtering mass as given by the linear calculation in
\citet{NB07} to the simulation results (Section~\ref{sec:linear}).
Finally, we summarize and discuss our
results in Section~\ref{sec:con}.


\section{The Simulations}
\label{sec:sims}

Our simulations assume the following cosmological parameters:
($\Omega_\Lambda$, $\Omega_{\rm M}$, $\Omega_b$, n, $\sigma_8$, $H_0$)
= (0.7, 0.3, 0.047, 1, 0.92, 70 km s$^{-1}$ Mpc$^{-1}$).  We use the
Eulerian adaptive mesh refinement (AMR) code Enzo, which is described
in greater detail in \citet{Bryan}, \citet{NB99} and \citet{Andrei}.
Our simulation volumes are initialized at $z_{\rm init}=99$, with
density perturbations drawn from the \citet{EH99} power spectrum. The
initial density fluctuations are assumed equal for the baryons and
dark matter, and the initial gas temperature is uniformly 110.7 K. The
parameters of our runs (identified in order of increasing mass
resolution as Very Low, Low, High, and Very High resolution) are shown
in Table~\ref{table}. In each case, we first run a low resolution dark
matter only run, in which the simulation volume $l_{\rm root}^3 $ is
resolved on a $128^3$ root grid.
 We include an additional $n_ {\rm ref}$ static levels
of refinement (listed in Table~\ref{table} for each run) inside this
central cube.  Furthermore, grid cells inside the central region are
allowed to dynamically refine so that the Jeans length is resolved by
at least 4 grid zones and no grid cell contains more than 4 times the
initial gas mass element.  Each additional grid level refines the mesh
length of the parent grid cell by a factor of 2.  We allow for a
maximum of 11 levels of refinement inside the refined central region,
granting us a spatial resolution of $l_{\rm root} / (128 \times
2^{11})$.  The dark matter particle mass for each run is shown in
Table~\ref{table}.  We also include the non-equilibrium reaction
network of nine chemical species (H, H$^+$, He, He$^+$, He$^{++}$,
$e^-$, H$_2$, H$_2^+$, H$^-$) using the algorithm of \citet{Ann97},
and initialized with post-recombination abundances from \citet{Ann96}.
Our analysis below is based on the central refined region; the low
resolution dark matter outside the refined region serves to provide
the necessary tidal forces to our refined region.

\begin{table}
 \caption{Main characteristics of the simulations.}
\label{table}
\begin{center}
\begin{tabular}{l c c c c }
\hline
{Simulation}& Box size& Static refinement & Inner refined & $M_{\rm dm}$  \\
{name} &     [Mpc$/$h]  &levels      & region [Mpc$/$h] & [M$_\odot$]  \\
\hline \hline
VLres  &1 & 2&0.25  &  747    \\
Lres   &1 & 3&0.25 & 93   \\
Hres  &0.35  & 3& 0.0875 &  4.0 \\
VHres   &0.25 &3 &0.0625 & 1.46   \\
\hline
\vspace{-0.7cm}
\end{tabular}
\end{center}
\end{table}

Each simulation includes two ionization scenarios: (1) no UV
background radiation (hereafter ``NoUV''), and (2) flash ionization at
$z=25$ (``Flash'') which instantaneously sets the gas temperature to
T=10,000 K and the hydrogen neutral fraction to $x_{\rm HI} = 10^{-3}$
throughout the simulation volume, but involves no heating thereafter.

We use the HOP algorithm \citep{EH98} on the DM particles to identify
DM halos. Since we are interested in the gas fraction, it is essential
to sum up both the DM mass and the gas mass over the same volume. We
obtain the total DM and baryonic mass of each halo by integrating the
densities over a sphere whose radius is the halo's virial
radius\footnote{We define the virial radius as the radius of the
spherical volume within which the mean density is $\Delta_c$ times the
critical density.  $\Delta_c$ is obtained through the fitting formula
in \citet{BN98} and is $\sim 178$ in an Einstein-DeSitter universe.
We find that the halo masses thus obtained generally agree within a
factor of two with masses obtained directly from the HOP algorithm
assuming a mean baryon fraction.}. Additionally, we discount halos
which have been substantially contaminated by the large
(low-resolution) DM particles outside of our refined region.
Specifically, we remove from our analysis halos with an average DM
particle mass greater than $\sim$ 130\% of the refined region's DM
mass resolution.

The simulations do not attempt to model global reionization. Instead
we focus on a short-lived UV background that can occur from a
localized star formation episode. Specifically, the Flash scenario
enabled \citet{Andrei} to compare their results to \citet{Oshea05}.
 and to check the importance of including additional dynamical effects of
prolonged photo-heating. For us, this is simply a good test case of a
time-variable Jeans mass at high redshift.

\section{The resolution dependence of the gas fraction in halos}\label{sec:res}

\begin{figure}
  \centering \includegraphics[width=84mm]{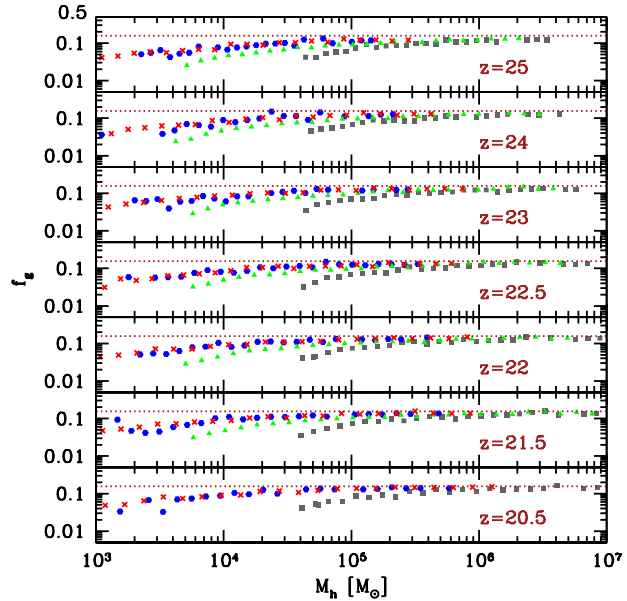}
\caption{NoUV case: The halo gas fraction $f_g$ as a function of the
halo mass $M_h$ in the simulations.  We consider the following
simulations runs: VLres (squares), Lres (triangles), Hres (crosses),
and VHres (circles). Note that we did not use the Lres data at
$z=20.5$ because of a numerical error.}
\label{fig:BIN}
\end{figure}

\begin{figure}
\centering
\includegraphics[width=84mm]{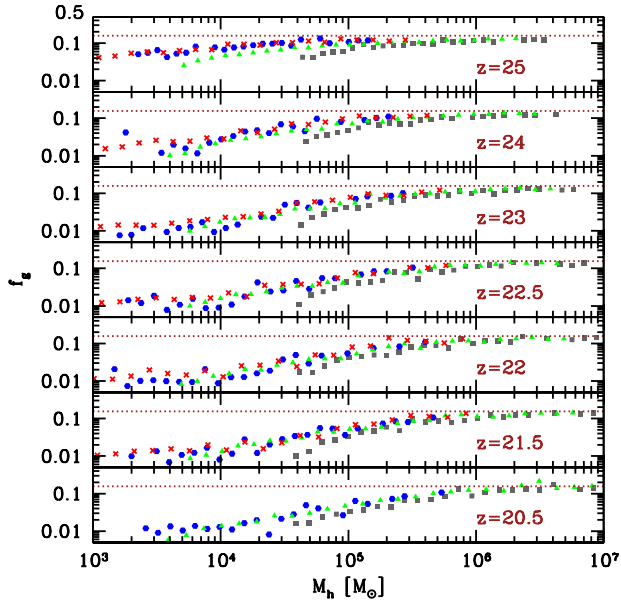}
\caption{Flash case: The halo gas fraction $f_g$ as a function of the
halo mass $M_h$ in the simulations. Same conventions as in
Fig.~\ref{fig:BIN}. Note that we did not use the Hres data at $z=20.5$
because of a numerical error.}
\label{fig:BIN_flash}
\end{figure}

Our simulation runs at four different resolutions (Table~\ref{table})
allow us to carefully test the dependence of the gas fraction
estimates on the resolution. We find that the halo gas fraction indeed
varies strongly with the resolution. This can be seen in
Figs.~\ref{fig:BIN} and \ref{fig:BIN_flash}, which show the gas
fraction in halos as a function of the halo mass. The gas fraction is
shown for each of our runs, averaged over bins of halo mass. These
figures show that the lower resolution runs underestimate the gas
fraction at a given halo mass, often substantially. However, our two
highest-resolution runs, Hres and VHres, agree in their gas fractions
over essentially their entire halo mass range, and above that range
(i.e., at $M \ga 10^6 M_\odot$) the two lower resolution runs agree. This
demonstrates numerical convergence and suggests that if we take our
highest-resolution run at each halo mass, then we obtain the correct
value of the mean halo gas fraction.

Getting a correct gas fraction in simulations is important in
calculating, for example, the galaxy abundance. Underestimation of
the gas fraction in halos can give incorrect results. The VLres
run was used by \citet{Andrei} and at the time was considered a
very high resolution run in the high-redshift regime. We find,
however, that our high resolution runs (i.e., Hres and VHres)
estimate the gas fraction in the NoUV case to be higher than for
the VLres run by a factor of $\sim 1.6$ for a halo mass of $10^5
M_\odot$. Note, that \citet{Andrei} focus on whether the gas has
cooled relative to other runs at the same resolution, not the
exact amount of gas inside each halo.  Furthermore, that study
focused on halos whose gas was capable of cooling via the
molecular hydrogen channel, corresponding to masses greater than
$\sim 5 \times 10^5 M_\odot$ at these redshifts, where the gas
fraction is beginning to converge to the higher-resolution values.

\begin{figure}
\centering
\includegraphics[width=84mm]{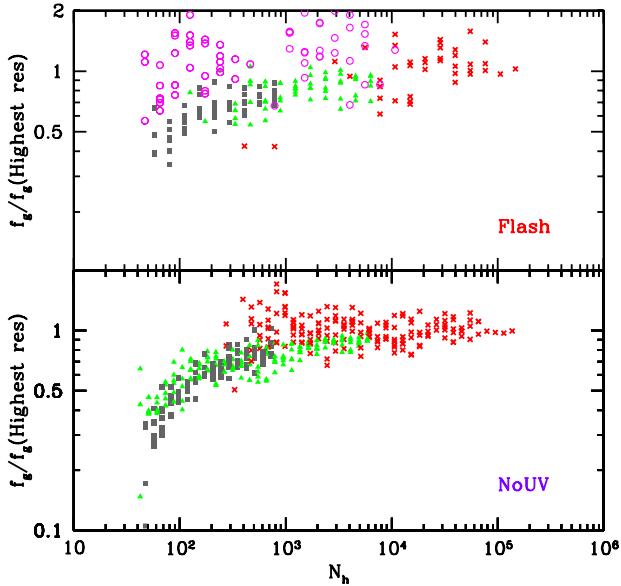}
\caption{The ratio between the halo gas fraction in one simulation and
the gas fraction in the highest-resolution run in the same mass bin,
as a function of the number of particles per halo ($N_h$) in the
lower-resolution run. We show these ratios from all the output
redshifts. We consider the NoUV (bottom panel) and Flash (upper panel)
cases, with symbols corresponding to the lower-resolution run using
the same conventions as in Fig.~\ref{fig:BIN}. For the Flash case we
only include halos for which $f_g>f_{g,min}=0.03$ in the
higher-resolution run (see text), and we separately show the results
from the excluded halos (open circles).}
\label{fig:all_fg}
\end{figure}

We can further quantify the resolution dependence of the halo gas
fraction. At each halo mass, we adopt the gas fraction as given by our
highest-resolution simulation available at that mass scale ($f_{g(\rm
  Highest-res)}$) to be the correct value, and compare to this the
result from all lower-resolution simulations that have halos of that
same mass at the same redshift.
We plot in Figure~\ref{fig:all_fg} these ratios, $f_g/f_{g(\rm
  Highest-res)}$, where each simulation at each output redshift is
compared to the highest-resolution run at the same halo mass and
redshift. In order to test the idea that the resolution effect
depends primarily on the halo mass resolution, these ratios are
shown as a function not of the halo mass but of the number of
particles in each halo ($N_h$) in the lower-resolution run. Note
that for this we divided our various runs into the same mass bins.
Now, if the halo gas fraction did not depend on resolution then we
would simply get unity. Instead, the figure shows that the lower
resolution runs underestimate the gas fractions in halos. In
Figure~\ref{fig:all_fg} we show together all the various redshift
outputs from all the different runs; all the redshift are
consistent with a single relation, i.e., they agree
quantitatively. Despite the scatter, there is a rather uniform
trend among all the points (excluding the open circles in the
Flash case -- see below), confirming the idea that the resolution
dependence is mainly an effect of halo mass resolution.

The gas fractions in some halos are less accurately determined in the
Flash case, for a simple reason. In this scenario the global heating
in the simulation evaporates the gas, i.e., the heating raises the
characteristic mass (see also section~\ref{sec:linear}). This is
easily seen in Fig.~\ref{fig:BIN_flash}, where we observe a declining
gas fraction with time in the low-mass range \citep[consistent
with][]{Haiman}.
Thus, the gas fraction in each halo is naturally more
sensitive to the halo's surroundings and to numerical errors when the
gas fraction is very low. Indeed, we find large scatter
below $f_g \sim 0.03$, which corresponds to low-mass halos (below
$M\sim 5\times 10^4$~M$_\odot$) at redshifts well after the UV flash.
We thus consider only the halos for which the gas fraction is larger
than the fiducial value $f_{g,min}=0.03$.\footnote{We make this
  separation only in this resolution analysis, while in section
  \ref{sec:linear} we calculate the scatter in each mass bin.} The
upper panel of Fig.~\ref{fig:all_fg} confirms the need for this
separation, showing that the Flash halos with $f_g>f_{g,min}$ follow a
similar trend as the halos in the NoUV case, while the excluded halos
with $f_g < f_{g,min}$ are inconsistent and show a much larger
scatter.

\begin{figure}
  \centering \includegraphics[width=84mm]{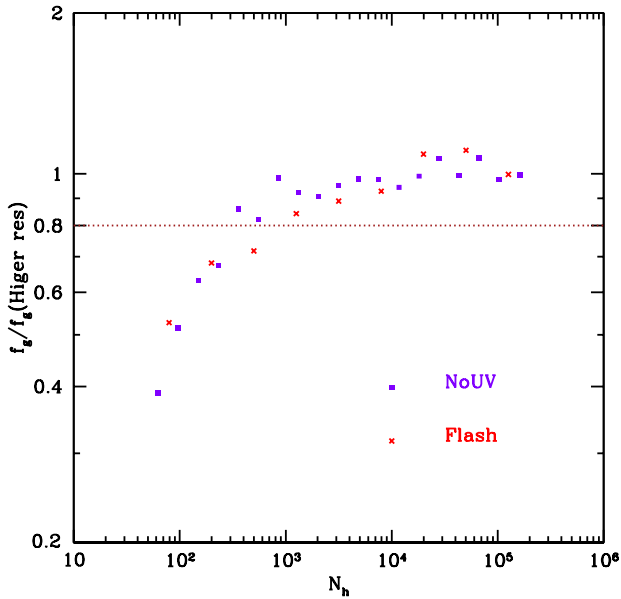}
\caption{The ratio between the halo gas fraction in one simulation and
the gas fraction in the highest-resolution run in the same mass bin,
as a function of the number of particles per halo ($N_h$) in the
lower-resolution run. We consider the NoUV case (squares) and the
Flash case (circles), where in each case all the points from the
various runs and output redshifts shown in Fig.~\ref{fig:all_fg} have
been condensed into a single set of bins. In the Flash case we include
only halos for which $f_g>f_{g,min}=0.03$ in the higher-resolution
run.}
\label{fig:all}
\end{figure}

To derive a useful result from Fig.~\ref{fig:all_fg}, we first
condense it efficiently by putting all the points together into a
single set of bins in $N_h$. Specifically, we show in
Fig.~\ref{fig:all} one set of bins for the NoUV case, and one for
the Flash case (including only the halos with $f_g>f_{g,min}$).
This figure shows a clear trend, consistent between the two cases,
of an artificially declining gas fraction in poorly resolved
halos. If we desire a reasonably accurate determination of the
mean halo gas fraction, with a systematic error of no more than
$20\%$ (i.e., $f_g/f_{g(\rm Highest-res)} \geq 0.8$), then at
least 500 particles per halo are required. For a $10\%$ error,
$\sim 2000$ particles are required per halo. On the other hand,
halos resolved into only 100 particles underestimate the enclosed
gas fraction by more than a factor of two. We adopt 500 as a
minimum number, and henceforth consider only halos for which $N_h
\geq 500$.

We note that, averaged over the output redshifts, the number of halos
that we extract from each simulation is 104, 1532, 2479, and 2304, for
the NoUV case and 102, 1641, 2601, and 324 for the Flash case, for the
VLres, Lres, Hres, and VHres simulation, respectively. We also note
that at redshift $20.5$ we did not use the data for the simulation
Lres for the NoUV case and Hres for the Flash case, due to numerical
problems in these particular outputs. It is important to point out
that we only performed this resolution test with Enzo. Although gas
properties in dense regions are found to agree well in AMR and SPH
cosmological simulations, different gravity solvers can lead to
non-negligible differences in the N-body mass function at low masses
\citep{Oshea05}. 
Specifically, it seems that the gravitational softening in Enzo causes
small halos to form somewhat late, but detailed resolution tests show
that the halo gas fractions at a given time are unaffected by
softening at least down to 100-particle halos \citep[see Figure~14
in][]{Oshea05b}.

\section{The characteristic mass at high redshift}
\label{sec:linear}

\subsection{Definition and relation to linear model}

In linear theory the filtering mass, first defined by \citet{cs},
describes the highest mass scale on which the baryon density
fluctuations are suppressed significantly compared to the dark matter
fluctuations. In \citet{NB07} we included the fact that the baryons
have smoother initial conditions than the dark matter
\citep[see][]{NB05} and found a lower value of the filtering mass (by
a factor of $3-10$, depending on the redshift). Following
\citet{NB07}, the filtering scale (specifically, the filtering
wavenumber $k_F$) is defined by expanding the ratio of baryonic to
total density fluctuations to first order in $k^2$:
\begin{equation}
\frac{\delta_\br}{\delta_\tot}=1-\frac{k^2}{k_F^2}+r_{\rm LSS}\ ,
\label{kf_btot}
\end{equation}
where $k$ is the wavenumber, and $\delta_\br$ and $\delta_\tot$ are
the baryonic and the total (i.e., including both baryons and dark
matter) density fluctuations, respectively. The parameter $r_{\rm
LSS}$ (a negative quantity) describes the relative difference between
$\delta_\br$ and $\delta_\tot$ on large scales \citep[for more details
see][]{NB07}. The filtering mass is defined from $k_F$ simply as:
\begin{equation}
M_F=\frac{4\pi}{3}\bar{\rho_0}\left(\frac{1
}{2}\frac{2\pi}{k_F}\right)^3\ ,
\label{Mf}
\end{equation}
where $\bar{\rho_0}$ is the mean matter density today. This relation
is one eighth of the definition in \citet{gnedin00} (based on a
non-standard definition of the Jeans mass used there).
 Following
\citet{NB07} we calculate the filtering mass for the cosmological
parameters assumed in this paper.

There is no apriori reason to think that the filtering mass can also
accurately describe properties of highly non-linear, virialized
objects. For halos, \citet{gnedin00} defined a characteristic mass
$M_c$ for which a halo contains half the mean cosmic baryon fraction
$f_b$. In his simulation he found the mean gas fraction in halos of a
given total mass $M$, and fitted the simulation results to the
following formula:
\begin{equation}
\label{f_g-alpha}
f_{g,\rm calc}= f_{\br,0} \bigg[1+\left(2^{\alpha/3}-1
\right)\left(\frac{M_c}{M}\right)^\alpha \bigg]^{-3/\alpha} \ ,
\end{equation}
where $f_{\br,0}$ is the gas fraction in the high-mass limit.  In this
function, a higher $\alpha$ causes a sharper transition between the
high-mass (constant $f_g$) limit and the low-mass limit (assumed to be
$f_g \propto M^3$). \citet{gnedin00} found a good fit for $\alpha =
1$, with a characteristic mass that in fact equaled the filtering mass
by his definition. By our definition in eq.~(\ref{Mf}), the claim from
\citet{gnedin00} is that $M_c=8\times M_F$.

Recently, \citet{Hoeft} and \citet{Okamoto} used higher resolution
simulations than in \citet{gnedin00} and showed that this claim is
incorrect in the low-redshift regime. They compared the characteristic
mass found in their simulations to the \citet{cs} filtering mass and
found that the two values diverge after reionization. Specifically,
they found that $\alpha = 2$ and $M_c(z=0)\sim 6.5\times
10^9$~M$_\odot/h$, which is much lower than the filtering mass (and
thus even lower compared to $8\times M_F$). We caution that unlike our
simple test cases, the heating in reionization simulations is complex
and inhomogeneous, and thus the filtering mass cannot be directly and
precisely defined and computed. Also note that the precise
quantitative results during and after reionization depend on the
thermal history of the gas which observationally is not well
constrained, though results at $z=0$ are more robust, as the gas in
overdensities around halos has had more time to ``forget'' the details
of its heating history.  Nevertheless, by simulating a large set of
thermal histories in a cosmological setting, \citet{MD08} were able to
draw a general conclusion that the characteristic mass towards the end
of reionization is likely in all cases to be close to the
atomic-cooling threshold of $\sim 10^8 M_{\odot}$.

\subsection{Simulation results}

\begin{figure}
\centering
\includegraphics[width=84mm]{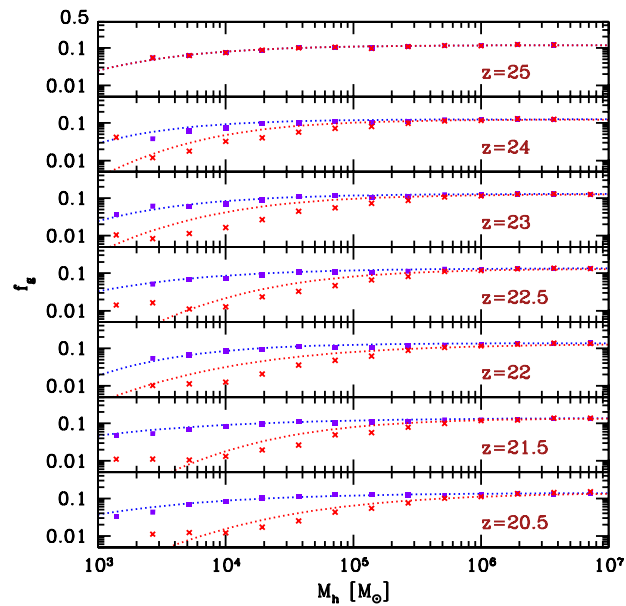}
\caption{The gas fraction in all the simulations vs.\ the halo mass at
  various redshifts. The simulation data have been binned by halo
  mass. We use only data points from halos with $N_h\geq 500$. We
  consider the Flash (crosses) and NoUV (filled squares) cases. Also
  shown are the best fits in each case in the form of
  eq.~(\ref{f_g-alpha}). We note that the bins show here are equally
  spaced for representations reasons, for the actual degrees of
  freedom see table \ref{table2}. }
\label{fig:BIN_mass1}
\end{figure}

In order to determine the characteristic mass from our simulations, we
put together the gas fraction measurements from all the simulation
runs at each redshift, but including only the well-resolved halos,
i.e., those with $N_h\geq 500$, as determined in
section~\ref{sec:res}. We fit the simulation results to
equation~(\ref{f_g-alpha}) with two free parameters, $M_c$ and
$\alpha$, taking $f_{\br,0}$ to be the average gas fraction in the
highest few mass bins\footnote{We note that we also tried to carry out
the analysis with $f_{\br,0}$ assumed to be the cosmic mean value.
This gave a substantially worse fit, with the parameters changed
significantly ($\sim 40\%$ and $\sim 10\%$ changes in $M_c$ and
$\alpha$, respectively).
 We also tried the approach of taking
$f_{\br,0}$ to be a free parameter (in the NoUV case). This made less
of a difference compared to taking $f_{\br,0}$ from the highest few
mass bins. We found a lower $M_c$ by $<10\%$, a lower $\alpha$ by
$<6\%$, and a fitted value for $f_{\br,0}$ higher by $<7\%$ than the
simple estimate we used in the text.}. We used a minimum-$\chi^2$
method to estimate the best fit to equation~(\ref{f_g-alpha}) and to
find the errors. To account for the numerical scatter we fitted this
equation to the binned data, adopting the standard deviation within
the mass bin as the uncertainty in each binned value. The best-fit
parameters for both the NoUV and the Flash scenarios along with their
1-$\sigma$ confidence limits are listed in Table~\ref{table2} (and
also shown in Fig.~\ref{fig:Mc}, which is discussed below). We also
compare the binned data to the corresponding best fits in the form of
eq.~(\ref{f_g-alpha}) in Figure~\ref{fig:BIN_mass1}.

\begin{table}
 \caption{ The best-fit parameters from equation~(\ref{f_g-alpha}).}
\label{table2}
\begin{center}
\begin{tabular}{l c c c c c}
\hline
Redshift & $M_c$               &  $\alpha$ & reduced $\chi^2$ &degrees
\\
          &  [$10^3$~M$_\odot$] &           & &of freedom\\
\hline \hline
 & & {\bf NoUV} & &\\
\hline \hline
$20.5$ & $4.52^{+1.45}_{-1.88}$&  $0.47^{+0.06}_{-0.07}$& $1.15$&32 \\
$21.5$ & $3.28^{+0.79}_{-0.99}$&  $0.41^{+0.03}_{-0.03}$ & $0.81$&52\\
$22$   & $5.65^{+0.17}_{-0.47}$&  $0.70^{+0.01}_{-0.03}$ & $1.3$&49\\
$22.5$ & $4.43^{+0.64}_{-0.93}$&  $0.53^{+0.04}_{0.05}$ & $0.54$&46\\
$23$   & $4.83^{+0.16}_{-0.16}$ &  $0.63^{+0.02}_{-0.02}$  & $1.4$&35\\
$24$   & $3.13^{+0.49}_{-0.47}$&  $0.75^{+0.21}_{-0.11}$  & $0.45$&33\\
$25$   & $4.12^{+0.69}_{-1.42}$ &  $0.64^{+0.06}_{-0.1}$& $0.21$&21\\
\hline \hline
 & & {\bf Flash} & &\\
\hline \hline
$20.5$ & $141.3^{+11.0}_{-12.8}$ &$0.49^{+0.02}_{-0.02}$ & $0.9$&31 \\
$21.5$ & $68.7^{+0.1}_{-1.41}$   &$0.64^{+0.01}_{-0.01}$& $1.6$&48\\
$22$ & $51.47^{+5.67}_{-5.25}$     &$0.48^{+0.02}_{-0.02}$ & $3.8$&54\\
$22.5$ & $55.99^{+0.27}_{-1.23}$   &$0.63^{+0.01}_{-0.01}$ & $1.58$&53\\
$23$  & $22.83^{+2.46}_{-2.43}$    &$0.62^{+0.03}_{-0.03}$ & $0.85$&36\\
$24$ & $16.2^{+3.4}_{-0.6}$     &$0.72^{+0.07}_{-0.07}$ & $0.96$&31\\
$25$ &  $4.12^{+0.69}_{-1.42}$     &$0.64^{+0.06}_{-0.1}$ & $0.17$&26\\
\hline
\vspace{-0.7cm}
\end{tabular}
\end{center}
\end{table}

\begin{figure}
  \centering \includegraphics[width=84mm]{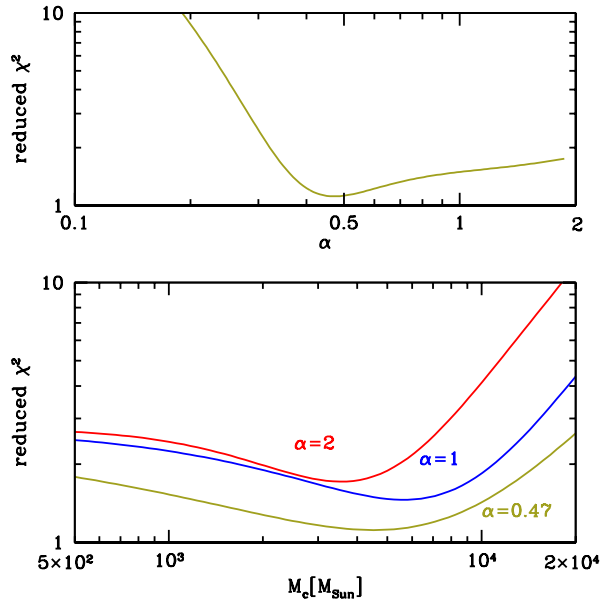}
\caption{The dependence of the reduced $\chi^2$ on the
parameters of the fit at $z=20.5$. There are 33 degrees of freedom in
this case. Bottom panel: dependence on $M_c$ obtained while fixing
$\alpha=2$, 1, or 0.47, from top to bottom respectively, where the
lowest value is the best-fit one (see table~\ref{table2}). Upper
panel: dependence on $\alpha$, fixing the best-fit value
$M_c=4.52\times 10^3$~M$_\odot$. }
\label{fig:chi2Mc-alpha}
\end{figure}

\begin{figure}
\centering
\includegraphics[width=84mm]{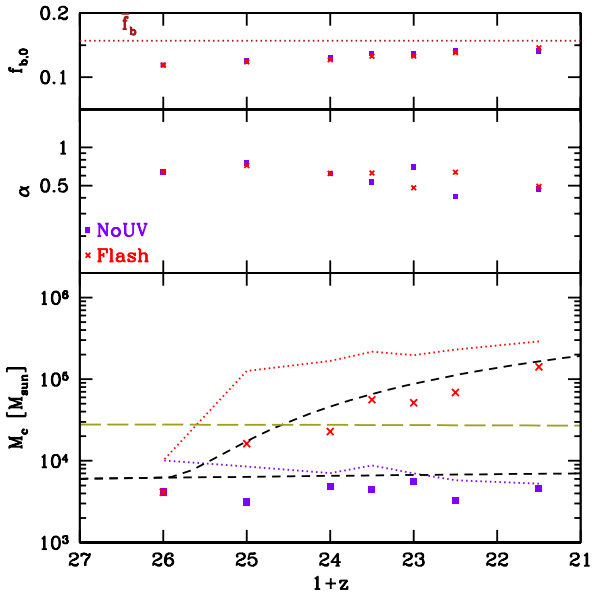}
\caption{ The parameters of the best fits in the form of
equation~\ref{f_g-alpha}; different panels show $M_c$, $\alpha$, and
$f_{\br,0}$. We consider the Flash (crosses) and NoUV (filled squares)
cases, where we fit equation~\ref{f_g-alpha} to all data points from
halos with $N_h\geq 500$. For comparison we also show the $M_c$ values
derived directly from interpolating the binned data, without assuming
a fitting function (dotted curves in the bottom panel; bottom: NoUV,
top: Flash scenario). In the bottom panel we also show the analytical
calculation according to \citet{NB07}, assuming the same initial
conditions as in the simulations (short-dashed curves; bottom: NoUV,
top: Flash). The full calculation assuming the true initial conditions
as in \citet{NB07} is also shown for the NoUV case (long-dashed
curve). We note that the NoUV and Flash cases mostly overlap in the
top panel, which also shows the cosmic mean baryon fraction
(horizontal dotted line).}
\label{fig:Mc}
\end{figure}

\citet{Hoeft} and \citet{Okamoto} found that the fits to their
simulations were consistent with $\alpha=2$. However, we find that our
fits yield $\alpha\sim 0.4 - 0.7$ (see Table~\ref{table2} and also
Fig.~\ref{fig:Mc} middle panel). In Figure~\ref{fig:chi2Mc-alpha} we
show the dependence of the reduced $\chi^2$ on the fit parameters
$M_c$ and $\alpha$ at $z=20.5$. This figure shows that a low $\alpha$
gives the best fit to equation~(\ref{f_g-alpha}), and suggests that
the characteristic mass scale found assuming $\alpha=2$ would be an
underestimate by about $15\%$ in this case.

Figure \ref{fig:Mc} shows the best fitted parameters at various
redshifts for $M_c$ and $\alpha$, and our value for $f_{\br,0}$, for
both the NoUV and Flash scenarios. The 1-$\sigma$ ($68\%$) confidence
regions are listed in table~\ref{table2}; since these statistical
errors are small, we do not show them in the figure.  Indeed,
particularly in the Flash case, the fits do not drop as quickly as the
data points do towards low halo masses, and so tend to systematically
underestimate the characteristic mass. Thus, to obtain a more
realistic estimate of the systematic error resulting from the form of
the fit, we also show $M_c$ as derived directly from the binned data
without a fit; in this case we simply found the maximum halo mass for
which $f_g=f_{\br,0}/2$ (interpolating inbetween the binned points).
We find that the systematic errors for the NoUV case (typical
difference of a factor of 1.5--2 between the fit and the no-fit
values) are substantially smaller than in the Flash case (typical
difference of a factor of 3), and they are much larger than the
statistical errors in both cases.

Figure~\ref{fig:Mc} also shows the analytical calculation from linear
theory of the filtering mass, as described in detail in \citet{NB07};
we make this calculation for both the NoUV and Flash scenarios,
assuming the same initial conditions as in the simulations. As
mentioned above, the simulation assumes equal baryon and dark matter
fluctuations at its $z_{\rm init}=99$ (as is commonly assumed in the
literature), while the correct baryonic initial conditions are
smoother (see below). Note that due to the simplicity of the Flash
scenario that we have implemented in the simulation, we can easily
incorporate it precisely within the analytical calculation. We also
directly tested the effect of radiative cooling, which is included in
the simulations but not the analytical model. We carried out an
additional Flash run with a resolution identical to that of the Hres
run, but where the radiative cooling was eliminated, leaving only
adiabatic cooling and Compton heating. We found that the gas fractions
(and thus the fitted characteristic mass) did not change
significantly, and thus verified that radiative cooling has a
negligible effect on our results.

We find that, given our systematic errors, the filtering mass from
linear theory is consistent with the characteristic mass fitted from
the simulations for both the NoUV and the Flash cases\footnote{ We
  note that equation~(\ref{f_g-alpha}) is successful in giving a
  reduced $\chi2$ of order unity in all cases except one (see
  Table~\ref{table2}).}. It is important to emphasize that in this
statement we are referring to our definition in equation~(\ref{Mf}),
which is one eighth of the original definition which \citet{gnedin00}
claimed was a good fit to the characteristic mass. In any case, we
conclude that at least in a particular redshift range ($z=20-25$) the
filtering mass provides a fairly good estimate to the characteristic
mass, either before stellar heating or in its initial stages. Since we
have not probed a larger range of redshifts, we cannot generalize this
conclusion. Also, the large systematic errors (particularly in the
Flash case) reduce the significance of the above conclusion.

In Figure~\ref{fig:Mc} we also show the filtering mass from the full
calculation of \citet{NB07} assuming the correct initial conditions
\citep[see figure 1 of][]{NB05}, in the NoUV case. The correct initial
conditions cannot be fully directly incorporated in a simulation
without starting at much higher redshifts than simulators are used
to. In particular, these initial conditions include the fact that at
$z=1200$ the baryons are still essentially uniform (on scales relevant
for galactic halos) due to their just ended strong coupling to the CMB
photons. Based on the agreement we have found between the linear
theory and the simulations for the case of the simulations' initial
conditions, we suggest that we can estimate the real characteristic
mass in the universe based on our analytical filtering mass
calculation with the true initial conditions.

There are several differences between our simulations' initial
conditions and the true ones. The initial conditions in the
simulations assumed a lower temperature (by $\sim 30\%$ at $z_{\rm
  init}=99$) than the exact calculation with Compton heating,
resulting in lower gas pressure and thus a lower filtering mass than
in the full analytical calculation. The assumption that the baryon
perturbations follow the dark matter at $z_{\rm init}=99$ creates
tendencies to both raise the filtering mass (since the baryon
fraction, and thus the gas pressure, is too high within perturbations)
and lower it (since the filtering mass reflectes the integrated effect
of pressure, and the integral is only begun at $z_{\rm init}=99$
instead of at $z=1200$). Also, we note that the best-known
cosmological parameters are slightly different from those in our
simulation, i.e., in our analytical calculation we use those of
\citet{wmap5}: ($\Omega_\Lambda$, $\Omega_{\rm M}$, $\Omega_b$, n,
$\sigma_8$, $H_0$) = (0.701, 0.299, 0.0478, 0.957, 0.82, 68.7 km
s$^{-1}$ Mpc$^{-1}$). However, changing the cosmological parameters
does not play an important role.  This is because the definition of
the filtering mass is independent of $\sigma_8$ which is the major
difference in the cosmological parameters. Thus, we conclude that the
true minimum mass needed for a halo to keep at least half of its
baryons, in the era before stellar heating (i.e., corresponding to the
NoUV case), is about $2.7\times 10^4$~M$_\odot$ at $z \sim 20-25$.

\begin{figure}
\centering
\includegraphics[width=84mm]{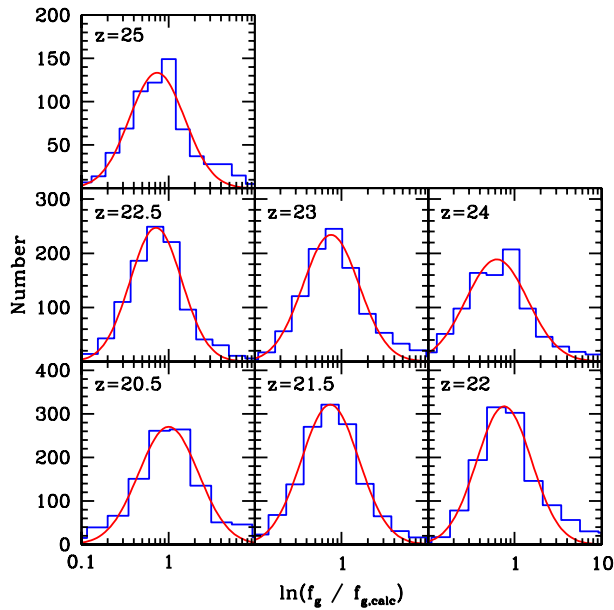}
\caption{The distribution of gas fractions with respect to the
prediction of equation~(\ref{f_g-alpha}). The histograms show the
binned data of $\ln(f_g/f_{g,\rm calc})$, where $f_{g,\rm calc}$ for
each halo mass is taken from eq.~(\ref{f_g-alpha}) assuming the
best-fit parameters as given in Table~\ref{table2}. We also show in
each case the best fit to a normal distribution (solid curves).}
\label{fig:pdf}
\end{figure}

Finally, we can also use our simulations to look beyond the tight
$f_g(M)$ relation of equation~(\ref{f_g-alpha}), and consider the
distribution of gas fractions for a given halo mass. \citet{gnedin00}
showed that this distribution in the simulation is well approximated
as a lognormal distribution.  Performing the same analysis for the
NoUV scenario, we find the same, as shown in Figure~\ref{fig:pdf}.
Since we have a limited number of halos, we collect all of our data at
each redshift and consider the ratio between the measured $f_g$ and
that predicted by equation~(\ref{f_g-alpha}) with the best-fit
parameters. Thus, we assume that this relative distribution does not
vary strongly with halo mass. Since it is easier to deal with a normal
distribution, we plot the distribution of $\ln (f_g/f_{g,\rm calc})$
and fit to it a normal distribution with mean $\mu$ and standard
deviation $\sigma$. We show the best-fit parameter values with their
1-$\sigma$ confidence ranges in Table~\ref{table3}.

\begin{table}
 \caption{The parameters of the best-fit normal distributions to the
 histograms shown in Figure~\ref{fig:pdf}.}
\label{table3}
\begin{center}
\begin{tabular}{l  c c c }
\hline
 redshift & $\mu$   & $\sigma$ &reduced $\chi^2$   \\
          &         &          &                   \\
\hline \hline
$20.5$ & $0.0003\pm 0.015$ &  $0.24\pm 0.01$&0.98 \\
$21.5$ & $-0.14\pm0.014$   &  $0.22\pm 0.01$&0.99\\
$22$   & $-0.12\pm 0.01$  & $0.22\pm 0.01$&0.97\\
$22.5$ & $-0.14\pm0.01$   & $0.21\pm0.006$&0.99\\
$23$   & $-0.12\pm0.01$    & $0.24\pm0.009$&0.98\\
$24$   & $-0.20\pm0.02$   & $0.24\pm0.01$&0.96\\
$25$   & $-0.13\pm0.02$   & $0.22\pm0.01$&0.93\\

\hline \vspace{-0.7cm}
\end{tabular}
\end{center}
\end{table}

\section{Conclusions}\label{sec:con}

We have used three-dimensional hydrodynamic simulations to investigate
the resolution requirements needed to determine correctly the gas
fraction in halos in the early universe. We considered both a NoUV
case with no stellar heating, and a Flash case with instantaneous
stellar heating.
 We found that the gas fraction in halos is
strongly dependent on the mass resolution of the simulation (see
Figs.~\ref{fig:BIN} and \ref{fig:BIN_flash}) both in the NoUV and
Flash cases. Using our multiple runs at various resolutions, we
demonstrated convergence in the estimated gas fractions over a wide
range of halo masses; thus we concluded that these estimates are
likely correct. Comparing these converged values to the results from
lower-resolution simulations, we showed (see Figs.~\ref{fig:all_fg}
and \ref{fig:all}) that halos that are poorly resolved (in terms of
the number of dark matter particles) yield artificially low gas
fractions. In particular, we concluded that to ensure a gas fraction
that is unbiased to $< 20\%$, there must be at least 500 particles in
each halo (and 2000 particles for $< 10\%$ bias). We showed that such
a simple condition is a consistent description of the resolution
dependence over the full range of redshifts and heating stages that we
investigated.

We found from the simulations the characteristic mass scale below
which a halo does not contain most of its baryons and can be
considered ``gas poor''. Specifically, we fitted eq.~(\ref{f_g-alpha})
to the data from the simulations with two parameters, the
characteristic mass $M_c$ and $\alpha$. We found that they are roughly
constant with redshift for the NoUV case (see Fig.~\ref{fig:BIN_mass1}
for the fit, also Fig.~\ref{fig:Mc} and table \ref{table2} for the
best fit parameters). We compared between the simulations'
characteristic mass and the analytical filtering mass, using the
simulations (with their particular initial conditions) as case
studies. We found that the $M_c$ estimations from the simulations are
consistent with the filtering mass from linear theory, according to
our definition which is one eighth of the definition that
\citet{gnedin00} claimed was a good fit to the characteristic
mass. Due to our limited redshift range and fairly large systematic
errors, we cannot be sure how general this consistency may be. We note
that this agreement between the simulations and the linear theory in
the NoUV and the Flash scenarios occurred in a regime where the
heating is simple and easily incorporated within linear theory. In
more complicated situations with inhomogeneous heating from
astrophysical sources, the filtering mass cannot be directly and
precisely computed, and this may explain apparent inconsistencies
between the theory and simulations
\citep{Hoeft,Okamoto}. Note that these authors used the
\citet{gnedin00} definition for the filtering mass, which is eight
times our definition.


The agreement between the linear calculation of the filtering mass and
the simulations suggests that using the correct initial conditions we
can calculate the true minimum mass in the real universe for which a
halo keeps most of its gas during its formation. We calculated the
filtering mass using the correct initial conditions starting from very
high redshift \citep[see,][]{NB05}, finding a characteristic mass of
$2.7\times 10^4$~M$_\odot$ assuming no stellar heating prior to $z
\sim 20$.  (see Fig.~\ref{fig:Mc}). We note that the filtering mass
for the Flash scenario with true initial conditions would have to be
corrected by approximately the same factor as $M_c$ in the NoUV case;
we did not show this curve in Fig.~\ref{fig:Mc} to avoid a busy
figure.

As noted in the introduction, minihalos can have important, observable
effects, in particular on the early stages of cosmic reionization. We
have taken an important step towards understanding the importance of
minihalos by establishing which ones contain substantial amounts of
gas. Also, since the minimum mass for molecular hydrogen cooling is
$\sim 10^5 M_{\odot}$, somewhat higher than the characteristic mass
for minihalos, we conclude that the gas fraction within the host halos
of the first stars could be slightly (though probably not greatly)
reduced compared to that of more massive halos; this effect is likely
missing or quantitatively inaccurate in many simulations of the first
stars, due to their inability to start with the correct initial
conditions at sufficiently high initial redshifts.



\section*{Acknowledgments}

We wish to thank Greg Bryan for helpful conversations and comments
on the paper. SN and RB acknowledge support by Israel Science
Foundation grant 629/05 and U.S. - Israel Binational Science
Foundation grant 2004386. RB is grateful for support from the
Moore Distinguished Scholar program at Caltech and the John Simon
Guggenheim Memorial Foundation. SN also acknowledges the support
of the John Bahcall Graduate Student Prize. Support for this work
was also partially provided by NASA through Hubble Fellowship
grant \#HF-01222.01 to AM, awarded by the Space Telescope Science
Institute, which is operated by the Association of Universities
for Research in Astronomy, Inc., for NASA, under contract NAS
5-26555.  We also acknowledge computational support from the
National Center for Supercomputing Applications.

\label{lastpage}


\begin{thebibliography}{09}


\bibitem[\protect\citeauthoryear{Abel et al.}{1997}]{abel97}
Abel T., Anninos P., Zhang Y., Norman M.~L., 1997, NewA, 2, 181

\bibitem[\protect\citeauthoryear{Abel, Bryan,
\& Norman}{2002}]{Abel02} Abel T., Bryan G.~L., Norman M.~L., 2002, Sci, 295, 93


\bibitem[\protect\citeauthoryear{Anninos
\& Norman}{1996}]{Ann96} Anninos P., Norman M.~L., 1996, ApJ, 460, 556


\bibitem[\protect\citeauthoryear{Anninos et
al.}{1997}]{Ann97} Anninos P., Zhang Y., Abel T., Norman
M.~L., 1997, NewA, 2, 209




\bibitem[\protect\citeauthoryear{Barkana \&
Loeb}{1999}]{BL99} Barkana R., Loeb A., 1999, ApJ, 523, 54

\bibitem[\protect\citeauthoryear{Barkana \& Loeb}{2001}]{rev}
Barkana, R., \& Loeb, A. 2001, Phys. Rep., 349, 125

\bibitem[\protect\citeauthoryear{Barkana \&
Loeb}{2002}]{bl02} Barkana R., Loeb A., 2002, ApJ, 578, 1

\bibitem[\protect\citeauthoryear{Barkana \& Loeb}{2005}]{BL05}
Barkana R., Loeb A., 2005, MNRAS,  363, L36

\bibitem[\protect\citeauthoryear{Benson et al.}{2002a}]{Benson02a}
Benson A.~J., Frenk C.~S., Lacey C.~G., Baugh C.~M., Cole S., 2002a, MNRAS,
333, 177


\bibitem[\protect\citeauthoryear{Benson et al.}{2002b}]{Benson02b}
Benson A.~J., Lacey C.~G., Baugh C.~M., Cole S., Frenk C.~S., 2002b, MNRAS,
333, 156


\bibitem[\protect\citeauthoryear{Bromm, Coppi, \&
    Larson}{2002}]{Bromm02} Bromm V., Coppi P.~S., Larson R.~B., 2002,
  ApJ, 564, 23



\bibitem[\protect\citeauthoryear{Bromm, Coppi, \&
    Larson}{1999}]{Bromm99} Bromm V., Coppi P.~S., Larson R.~B., 1999,
  ApJ, 527, L5


\bibitem[\protect\citeauthoryear{Bromm \& Loeb}{2004}]{BL04} Bromm V.,
  Loeb A., 2004, NewA, 9, 353


\bibitem[\protect\citeauthoryear{Bullock, Kravtsov,
\& Weinberg}{2000}]{Bullock} Bullock J.~S., Kravtsov A.~V., Weinberg D.~H., 2000, ApJ, 539, 517


\bibitem[\protect\citeauthoryear{Bryan}{1999}]{Bryan} Bryan G.~L.,
  1999, CoScE, 1, 46

\bibitem[\protect\citeauthoryear{Bryan
\& Norman}{1998}]{BN98} Bryan G.~L., Norman M.~L., 1998, ApJ, 495, 80

\bibitem[\protect\citeauthoryear{Cen}{2003}]{Cen} Cen R.,
2003, ApJ, 591, L5

\bibitem[\protect\citeauthoryear{Dijkstra et al}{2004}]{Dijkstra} Dijkstra M., Haiman Z., Rees M.~J.,
Weinberg D.~H., 2004, ApJ, 601, 666

\bibitem[\protect\citeauthoryear{Dunkley et
al.}{2008}]{Dunkley08} Dunkley J., et al., 2008, arXiv,
arXiv:0811.4280

\bibitem[\protect\citeauthoryear{Efstathiou}{1992}]{Efstathiou}
Efstathiou G., 1992, MNRAS, 256, 43P

\bibitem[\protect\citeauthoryear{Eisenstein
\& Hu}{1999}]{EH99} Eisenstein D.~J., Hu W., 1999, ApJ, 511, 5

\bibitem[\protect\citeauthoryear{Eisenstein
\& Hut}{1998}]{EH98} Eisenstein D.~J., Hut P., 1998, ApJ, 498, 137



\bibitem[\protect\citeauthoryear{Fuller}{2000}]{fuller}
Fuller T.~M., Couchman H.~M.~P., 2000, ApJ, 544, 6

\bibitem[\protect\citeauthoryear{Furlanetto
\& Oh}{2006}]{Furlanetto06} Furlanetto S.~R., Oh S.~P., 2006, ApJ, 652, 849

\bibitem[\protect\citeauthoryear{Gnedin \& Hui}{1998}]{cs} Gnedin,
  N.~Y. \& Hui, L. 2004, MNRAS, 296.

\bibitem[\protect\citeauthoryear{Gnedin}{2000}]{gnedin00} Gnedin
  N.~Y., 2000, ApJ, 542, 535.

\bibitem[\protect\citeauthoryear{Haiman \& Holder}{2003}]{HH03} Haiman
  Z., Holder G.~P., 2003, ApJ, 595, 1

\bibitem[\protect\citeauthoryear{Haiman, Rees \&
    Loeb}{1997}]{Haiman}{Haiman}, Z., {Rees}, M.~J. \& {Loeb}, 1997,
  ApJ, 476.

\bibitem[\protect\citeauthoryear{Helly et al.}{2003}]{Helly} Helly
  J.~C., Cole S., Frenk C.~S., Baugh C.~M., Benson A., Lacey C.,
  Pearce F.~R., 2003, MNRAS, 338, 913


\bibitem[\protect\citeauthoryear{Hoeft et al.}{2006}]{Hoeft} Hoeft M.,
  Yepes G., Gottl{\"o}ber S., Springel V., 2006, MNRAS, 371, 401

\bibitem[\protect\citeauthoryear{Hui
\& Gnedin}{1997}]{HG97} Hui L., Gnedin N.~Y., 1997, MNRAS, 292, 27

\bibitem[\protect\citeauthoryear{Iliev et al.}{2003}]{iliev2} Iliev, I.~T.,
Scannapieco, E., Martel, H., \& Shapiro, P.~R.\ 2003, MNRAS, 341, 81

\bibitem[\protect\citeauthoryear{Iliev et al.}{2005}]{iss05}
Iliev I.~T., Scannapieco E., Shapiro P.~R., 2005, ApJ, 624, 491

\bibitem[\protect\citeauthoryear{Jeans}{1928}]{jeans}Jeans, J.~H.
  Cambridge University press, (1928).

\bibitem[\protect\citeauthoryear{Kitayama
\& Ikeuchi}{2000}]{KI00} Kitayama T., Ikeuchi S., 2000, ApJ, 529, 615

\bibitem[\protect\citeauthoryear{Kuhlen, Madau,
\& Montgomery}{2006}]{Kuhlen} Kuhlen M., Madau P., Montgomery R., 2006, ApJ, 637, L1

\bibitem[\protect\citeauthoryear{Ma \& Bertschinger}{1995}]{Ma} Ma C.,
  Bertschinger E., 1995, ApJ, 455, 7

\bibitem[\protect\citeauthoryear{McQuinn et al.}{2007}]{mcquinn07}
McQuinn M., Lidz A., Zahn O., Dutta S., Hernquist L., Zaldarriaga M.,
2007, MNRAS, 377, 1043

\bibitem[\protect\citeauthoryear{Mesinger, Bryan \&
    Haiman}{2006}]{Andrei}Mesinger A., Bryan G., \& Haiman Z. 2006,
ApJ, 648, 835

\bibitem[\protect\citeauthoryear{Mesinger, Bryan,
\& Haiman}{2008}]{MBH08} Mesinger A., Bryan G.~L., Haiman Z., 2008, arXiv, arXiv:0812.2479 


\bibitem[\protect\citeauthoryear{Mesinger
\& Dijkstra}{2008}]{MD08} Mesinger A., Dijkstra M., 2008, MNRAS, 390,
1071  


\bibitem[\protect\citeauthoryear{Mesinger, Johnson,
\& Haiman}{2006}]{MJH06} Mesinger A., Johnson B.~D., Haiman Z., 2006, ApJ, 637, 80 


\bibitem[\protect\citeauthoryear{Nagashima, Gouda, \&
    Sugiura}{1999}]{Nag} Nagashima M., Gouda N., Sugiura N., 1999,
  MNRAS, 305, 449


\bibitem[\protect\citeauthoryear{Naoz \& Barkana}{2005}]{NB05} Naoz
  S., Barkana R., 2005, MNRAS, 362, 1047


\bibitem[\protect\citeauthoryear{Naoz, Noter, \& Barkana}{2006}]{NNB}
  Naoz S., Noter S., Barkana R., 2006, MNRAS, 373, L98


\bibitem[\protect\citeauthoryear{Naoz \& Barkana}{2007}]{NB07} Naoz
  S., Barkana R., 2007, MNRAS, 377, 667

\bibitem[\protect\citeauthoryear{Naoz
\& Barkana}{2008}]{NB08} Naoz S., Barkana R., 2008, MNRAS, 385, L63

\bibitem[\protect\citeauthoryear{Navarro, Frenk, \& White}{1997}]{NFW}
  Navarro J.~F., Frenk C.~S., White S.~D.~M., 1997, ApJ, 490, 493

\bibitem[\protect\citeauthoryear{Norman \& Bryan}{1999}]{NB99} Norman
  M.~L., Bryan G.~L., 1999, ASSL, 240, 19

\bibitem[\protect\citeauthoryear{Okamoto, Gao, \&
    Theuns}{2008}]{Okamoto} Okamoto T., Gao L., Theuns T., 2008,
  arXiv, 806, arXiv:0806.0378

\bibitem[\protect\citeauthoryear{O'Shea et al.}{2005a}]{Oshea05} O'Shea
  B.~W., Abel T., Whalen D., Norman M.~L., 2005, ApJ, 628, L5


\bibitem[\protect\citeauthoryear{O'Shea et al.}{2005b}]{Oshea05b} O'Shea
  B.~W., Abel T., Whalen D., Norman M.~L., 2005, ApJS, 169, 1





\bibitem[\protect\citeauthoryear{Peebles}{1980}]{Peebles} Peebles
  P.~J.~E., 1980, The Large-Scale Structure of the Universe, Princeton
  Univ. Press, Princeton


\bibitem[\protect\citeauthoryear{Press \& Schechter}{1974}]{Press}
Press W.~H., Schechter P., 1974, ApJ, 187, 425

\bibitem[\protect\citeauthoryear{Quinn, Katz, \&
    Efstathiou}{1996}]{Quinn} Quinn T., Katz N., Efstathiou G., 1996,
  MNRAS, 278, L49


\bibitem[\protect\citeauthoryear{Sheth \& Tormen}{2001}]{Sheth}
Sheth R.~K., Mo H.~J., Tormen G., 2001 ,MNRAS, 323, 1


\bibitem[\protect\citeauthoryear{Sheth \& Tormen}{2002}]{Sheth02}
Sheth R.~K., Tormen G., 2002, MNRAS, 329, 61

\bibitem[\protect\citeauthoryear{Shapiro et
al.}{2006}]{Shapiro06} Shapiro P.~R., Ahn K., Alvarez M.~A.,
Iliev I.~T., Martel H., Ryu D., 2006, ApJ, 646, 681

\bibitem[\protect\citeauthoryear{Shapiro, Giroux,
\& Babul}{1994}]{Shapiro94} Shapiro P.~R., Giroux M.~L., Babul A., 1994, ApJ, 427, 25





\bibitem[\protect\citeauthoryear{Somerville}{2002}]{Somerville}
  Somerville R.~S., 2002, ApJ, 572, L23


\bibitem[\protect\citeauthoryear{Springel \& Hernquist}{2003}]{SH03}
  Springel V., Hernquist L., 2003, MNRAS, 339, 312

\bibitem[\protect\citeauthoryear{Spergel et
al.}{2007}]{wmap5} Spergel D.~N., et al., 2007, ApJS, 170,
377



\bibitem[\protect\citeauthoryear{Tegmark}{1997}]{th2}
Tegmark M., et al., 1997, ApJ, 474, L77

\bibitem[\protect\citeauthoryear{Thoul
\& Weinberg}{1996}]{TW96} Thoul A.~A., Weinberg D.~H., 1996, ApJ, 465, 608


\bibitem[\protect\citeauthoryear{Trenti
\& Stiavelli}{2009}]{Trenti} Trenti M., Stiavelli M., 2009, arXiv,
arXiv:0901.0711



\bibitem[\protect\citeauthoryear{Wyithe \& Loeb}{2003}]{WL03} Wyithe
  J.~S.~B., Loeb A., 2003, ApJ, 588, L69


\bibitem[\protect\citeauthoryear{Yoshida}{2005}]{Yoshida}
Yoshida N., 2005, PThPS, 158, 117


\end{thebibliography}
\end{document}